\documentclass{article}
\usepackage{PRIMEarxiv}

\usepackage[table]{xcolor}
\usepackage[bookmarks=false]{hyperref}
\usepackage[utf8x]{inputenc}
\usepackage[T1]{fontenc}
\usepackage{amsmath}
\usepackage{dblfloatfix}
\usepackage[subrefformat=parens,labelformat=parens]{subfig}
\usepackage[justification=centering,skip=2pt,font={small}]{caption}
\usepackage{graphicx}
\usepackage{diagbox}
\usepackage{multirow}
\usepackage{comment}
\usepackage{adjustbox}
\usepackage{soul}
\usepackage{array}
\usepackage{makecell}
\usepackage{xspace}
\newcommand{\ie}[0]{{\em i.e.},\xspace}

\title{SToN: A New Fundamental Trade-off for Distributed Data Storage Systems}

\author{
  Bastien Confais \\
  LS2N, UMR 6004\\
  Polytech Nantes \\
  rue Christian Pauc, 44306 Nantes, France \\
   \And
  Şuayb Ş. Arslan \\
  Department of Computer Engineering\\
  MEF University\\
  Maslak,  34099 Istanbul, Turkey
   \And
  Beno\^it Parrein \\
  Nantes Université, LS2N, UMR 6004\\
  Polytech Nantes \\
  rue Christian Pauc, 44306 Nantes, France
}

\definecolor{y_color}{RGB}{255,245,206}
\definecolor{g_color}{RGB}{232,242,161}
\definecolor{b_color}{RGB}{222,230,239}
\definecolor{r_color}{RGB}{220,0,0}
\definecolor{o_color}{RGB}{255,128,0}

\newcolumntype{R}[2]{%
    >{\adjustbox{angle=#1,lap=\width-(#2)}\bgroup}%
    l%
    <{\egroup}%
}
\newcommand*\rot{\multicolumn{1}{R{60}{2em}}}%

\definecolor{review_color}{RGB}{44,136,160}

\begin{document}
\maketitle
\begin{abstract}
Locating data efficiently is a key process in every distributed data storage solution and particularly those deployed in multi-site environments, such as found in Cloud and Fog computing.
Nevertheless, the existing protocols dedicated to this task are not compatible with the requirements of the infrastructures that underlie such computing paradigms. In this paper, we initially review three fundamental mechanisms from which the existing protocols are used to locate data. We will demonstrate that these mechanisms all face the same set of limitations and seem to have a trade-off in three distinct domains of interest, namely, 
%traditional approaches share the same concepts of information-centric network approaches.
i) the scalability, ii) the ability to deal with the network topology changes
and iii) the constraints on the data naming process. After laying out our motivation and identifying the related trade-offs in existing systems, we finally propose a conjecture (and provide a proof for this conjecture) stating that these three properties cannot be met simultaneously, which we believe is a new fundamental trade-off the distributed storage systems using the three fundamental mechanisms have to face.
We conclude by discussing some of the implications of this novel result. 
\end{abstract}


\section{Introduction}

Fog computing infrastructures have been proposed as an alternative to cloud computing to provide computation abilities with minimal latency guarantees which makes it particularly suited for the Internet of Things (IoT) era~\cite{Bonomi:2012:FCR:2342509.2342513}.
It is clear that low latency response times for heavy-load computing tasks should come with an efficient data storage system backend.
In particular, this infrastructure is to be spread geographically in several sites for endurance and reliability. On the other hand,
designing a scalable storage solution that constrains the network traffic while at the same time places the data at an appropriate location to provide the lowest access times possible is a major challenge~\cite{7033259,8014365}. As of today, no consensus exists to our knowledge on the means and ways to be able to address this issue. One of the important aspects of a distributed data storage solution is 
the ability to locate data in an efficient manner.
In a previous work~\cite{confais:hal-01587459}, it has been argued that an ideal storage solution should possess five important properties: \textit{(i)}~data locality,
\textit{(ii)}~network containment, \textit{(iii)}~disconnected mode of operation, \textit{(iv)} mobility and \textit{(v)} scalability.

\begin{figure}[!t]
 \centering 
 \includegraphics[width=\columnwidth]{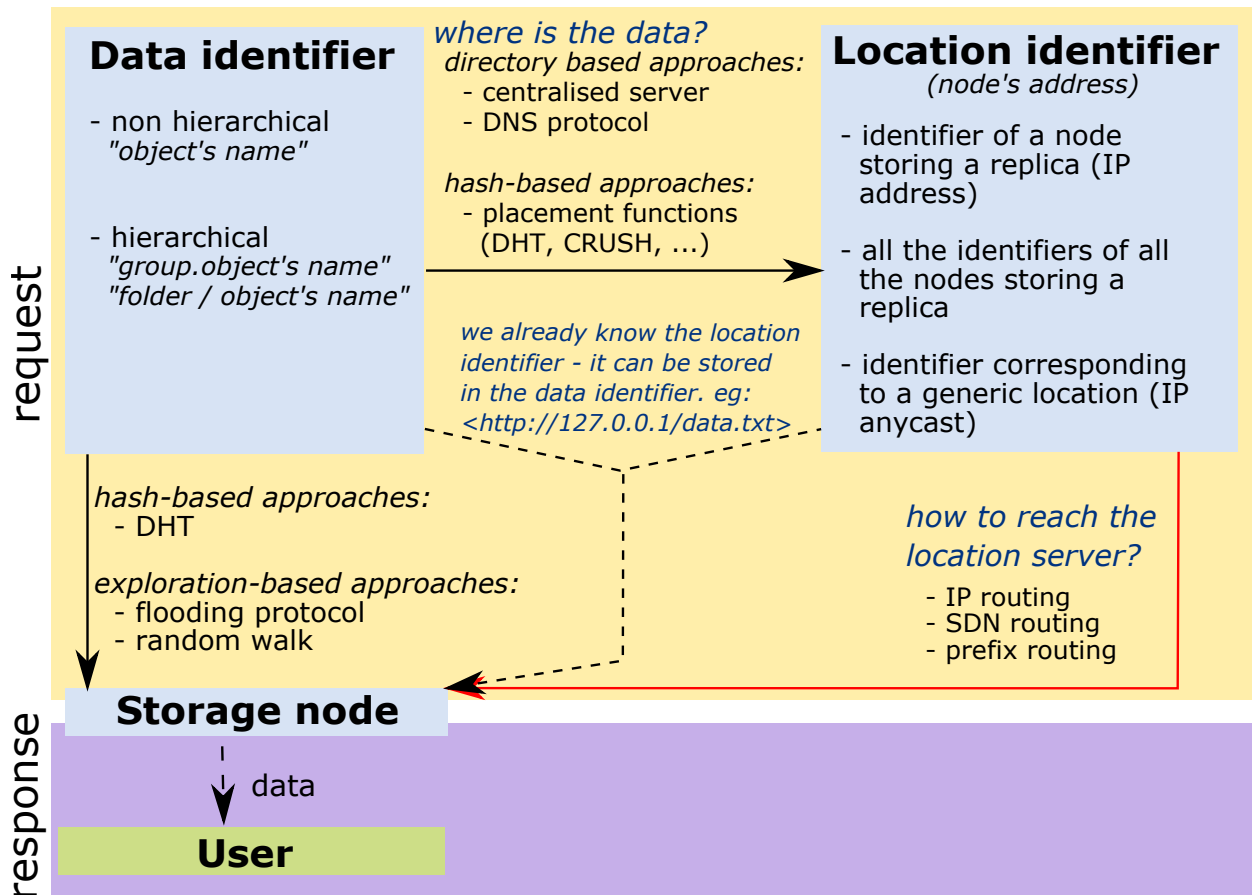}
 \caption{A generic process including request and response to access a data object in a distributed environment.}
 \label{fig:classification}
 \vspace{-.5cm}
\end{figure}

The main role of a data location protocol is to access the target server storing a specific piece of data through the use of a data identifier known by the user (``Data Identifier'' block on Figure~\ref{fig:classification}). Depending on the application, this can be an \textit{Uniform Resource Identifier} (URI), a filename, or any string that can be uniquely associated with each piece of data. 
Many protocols exist for locating data from the data identifier. Protocols like a Distributed Hash Table (DHT) or approaches like Information-Centric Networks (ICN) take the identifier and route the request directly to the node storing the piece of data (arrow between the ``Data identifier'' block and the storage node on Figure~\ref{fig:classification}), while others primarily identify a server address before routing the request through an IP network (arrow between the ``Data identifer'' block and the ``Location identifier'' block on Figure~\ref{fig:classification} and the arrow between the ``Location identifier'' block and the storage node).
%associate the server address to the 
%before routing the request.
%that associate from a data identifier, the server address, or the node storing the specific piece of data.
 Figure~\ref{fig:classification} illustrates these two scenarios in a generic environment with protocols locating data directly from the Data identifier and other that needs to use an intermediate location identifier.

%, they take as input the data identifier or the data name and either 
%produce the identifier of the node storing the data or directly transmit the request to this node.
For instance, protocols such as DNS act like a mapper between the identifier of the data and the identifier of the node storing the data.
Similar functionality can be observed in protocols such as Distributed Hash Table (DHT) or an Information Centric Network (ICN) routing.
Finally, the last category consists of protocols that take as input both the data identifier and the location identifier.

All of these protocols can be implemented in different ways and yet they all have to satisfy at most two of the three properties dictated by the Brewer's theorem (CAP theorem)~\cite{Gilbert:2002:BCF:564585.564601}: Consistency, Availability and a failure tolerance mechanism against network Partitioning.

In this study, the set of contributions can be listed as follows: 
\begin{itemize}
    \item We clearly demonstrate the limitations of the existing protocols for three mechanisms used to locate the data in a distributed environment;
    \item For the existing protocols considered, we establish an important trade-off that no distributed data storage solution can be Scalable (S), Tolerant to the Dynamic Network topology changes (To) and allow No Constraints (cannot impose any limitations) on the data Naming (N) or on the data placement;
    \item We finally provide a formal proof for this conjecture that we believe characterizes a new fundamental trade-off in distributed data storage systems using the three fundamental mechanisms described earlier. 
\end{itemize}

The provided proof applies to the existing protocols we could have identified in the literature. However, it is not general proof, rather proof that covers the three types (Directory-based, Hash-based, Exploration-based) of protocols which we believe are the only protocols presently used by all existing systems.
%In other words, the proof may not work for the systems we have not identified.}

The rest of the article is organised as follows. 
In Section~\ref{sec:properties}, we present the properties we expect from an ideal data location protocol in a Fog computing environment.
Then, in Section~\ref{sec:comparative}, we propose to classify the data location protocols in three different categories. 
For each category, we present some existing protocols and show why they do not meet the properties presented in the previous section. %how they are not adapted to a context of Fog Computing.
We then propose a fundamental conjecture that every data access protocol of a given distributed data storage system 
has a trade-off between  i) scalability, ii) tolerance to physical network topology changes and iii) constraints on the data placement and naming.
We finally formulate a proof of this conjecture in Section~\ref{sec:proof}, and point out limitations and implications of this new result in Section~\ref{sec:limitation}. We finally conclude our paper in Section~\ref{sec:conclusions}.

%%produce
%to the best of our knowledge, no data location protocol has been specifically proposed for Fog infrastructures.
%Nevertheless, many protocols have been proposed for accessing data in distributed environment in general.

\section{Properties of an ideal data location protocol}\label{sec:properties}

We initially present the properties we desire in a data location protocol designed for a multi-site environment,
and in particular for a Fog computing environment.

\paragraph{\textbf{Scalability (S)}}
Scalability is the ability of a process, a network, or a software program to grow and meet the increased demand without any observable performance degradation.
Two conclusions can be drawn from scalability: There is no node contacted in each request that would create a system bottleneck and there is no database that is growing boundlessly which stores a record for each piece of data stored in the system.

%Scalability is the ability of a process, a network, or a software program to grow and to meet the increased demand without any performance degradation. 
%For us it means two things, there is no node creating a Single Point of Failure in the network and there is no explosion
%of storage size in terms of location records stored.

To our point of view, the scalability is an important performance measure because a protocol adapted in a Fog computing environment should be able to work with a large number of nodes geographically spread and store large volumes of data at the same time. 
Jogalekar~\textit{et al.}~\cite{862209} proposed to measure the scalability of a distributed system through Quality of Service (QoS) metrics. We instead propose in this article to measure scalability using:

\begin{enumerate}
\item the size of the table used to store the data location;
\item the number of nodes storing such tables;
\item the number of requests to determine the location of data;
\item the number of network links used during this process.
\end{enumerate}

To our point of view, a system is scalable as long as the execution of a request does not require to contact all the nodes in the network or it does not require each node to store a record per data fragment.

\paragraph{\textbf{Tolerance to Dynamicity of the Network (To)}}
The Tolerance (To) to the physical network topology changes is the ability for a system to \textbf{automatically} reconfigure when the network topology is modified, i.e.,
some network links or nodes are added or removed, having its location identifier (\ie its IP address) changed. 
In other words, changing the topology does not require any additional action of the system's administrator.
Note that such changes in the physical network topology also require to update the network overlay to reflect the newly added or removed nodes.

The difference between To and the Partition tolerance (P) in the conventional CAP context is that in To any network topology change is permanent and not the consequence of a failure that would likely be fixed in the near future. This property can be evaluated/quantified with the network traffic needed to reconfigure the data storage solution and the associated number of reconfiguration steps.

%The tolerance to Dynamicity of the Network is the ability for a system to automatically reconfigure its nodes when the network topology change. That some
%links and nodes are added or removed.

\paragraph{\textbf{No Constraints on data Naming/Placement (N)}}

The absence of any constraint on data naming
is the ability for the user to choose the name (data identifier) used to access the contents of the data.
In the same way, the absence of constraint on the data placement enables the user to choose the node on which they want to store
their data. This enables the user to make a choice regarding the
policy in terms of privacy, durability or access rights proposed by different servers. Moreover, it provides a "stable" name that does not change regarding the placement of the data fragments if the data has to move for technical reasons, without the direct consent of the user.

%Finally, the absence of constraint on naming or placement is the ability for the user to choose the node on which they want to store
%their data but also the name used to access them.

%Finally, the constraints on the name of the data or on the placement is necessary to enable user to choose the identifier of the pieces of data they store
%and to place it where it is needed to reduce the access times.

\textbf{Notation:} In the following, we use $n$ as the number of nodes in the network and $d$ to represent the number of data chunks stored.
We also denote $diameter(n)$ to be the path with the highest number of hops between any two hosts in the network. 
If the topology is well-balanced, the equality $diameter(n)=log(n)$ usually holds but if all the nodes are interconnected within a chain topology, we shall have $diameter(n)=n$.

\section{A comparative study of the existing protocols}\label{sec:comparative}
In previous sections, we have presented the properties that an ideal data location protocol must possess. Let us now
explore how the existing solutions behave within the context of these properties, namely the scalability, ability to tolerate physical network topology changes, and the constraints on the data placement/naming.

To simplify the presentation, we propose a classification of the existing protocols that are used to locate data  into three distinct categories:
\begin{enumerate}
\item Protocols based on a directory which can be implemented in centralized or distributed fashion;
\item Protocols based on some form of computation;
\item Protocols based on an exploration of the network.
\end{enumerate}

As we shall see, this crude classification enable us to simplify the presentation of the concepts used by different protocols. We also believe that any other existing protocol can be classified under either one of these categories.

%We strongly believe that no other category currently exists but it is something %that needs to be formally demonstrated.

\subsection{Directory-based protocols}

This category of protocols relies on a database which can r{either} be centralised or distributed. These databases store the link 
between the data identifier and the location of a replica (the original content of the data).

\subsubsection{Central server}

The first approach is to deploy this database on a central server which is then requested each time a client needs to locate the data.
This approach is used in many distributed data storage solutions such as
High Performance Computing (HPC) (for instance, in PVFS~\cite{Carns:2000:PPF:1268379.1268407}, Lustre~\cite{donovan2003lustre})
or I/O intensive Scale-Out NAS (for instance in RozoFS~\cite{pertin:hal-01149847}).

In all of these solutions, clients primarily inquire about a central server to determine the right node storing the data fragment before any attempt is made to reach out that node. The main advantage is that adding or removing a storage node does not impact the central server and each data fragment can be localized in only a single request.

The main drawback of this approach is its centralized nature that degrades the scalability.
Using a centralized server cannot handle the burden of the increased volume of simultaneous requests and store the location
of the entire data in the system without noticeable performance degradation. 

\subsubsection{Domain Name System (DNS) protocol}

\begin{figure}[!t]
 \centering 
 \includegraphics[width=.9\columnwidth]{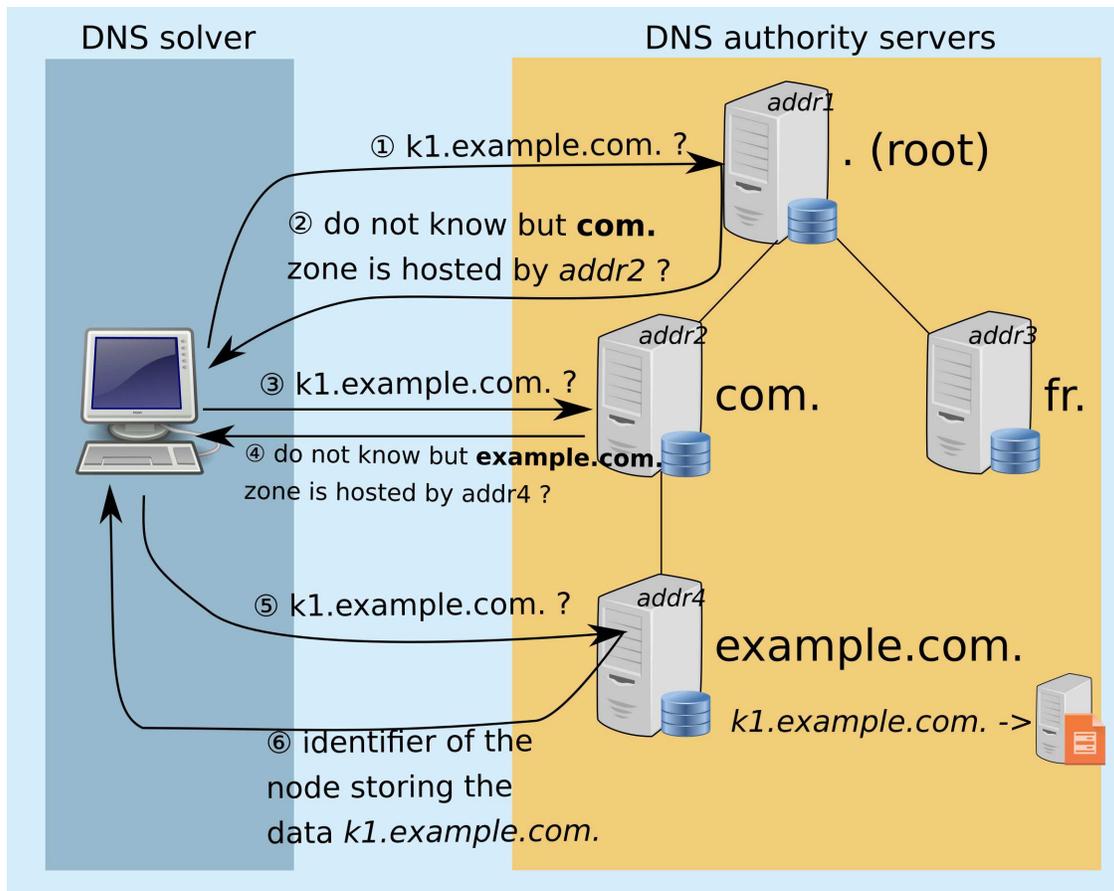}
 \caption{Iterative process of the DNS look-up process.}
 \label{fig:dns_resolution}
\end{figure}

In order to remove the bottleneck due to using a centralized server, the Domain Name System (DNS) protocol was proposed. DNS is a distributed directory~\cite{rfc1034} designed to provide scalability. It relies not only on a hierarchical namespace of identifiers but also on a hierarchy of servers.
Each server stores the data fragment locations with an identifier ending with a specific suffix. When the client initiates a request, the root server replies and returns the address of the next server managing the data's identifiers matching a given suffix.
The client then requests from this new server which can either return the data location directly or the address of another server managing the data's identifiers of a more specific suffix. For instance, in Figure~\ref{fig:dns_resolution}, the client initially requests the servers responsible for the root zone which only returns the address of the server responsible for the \texttt{com.}, then, it requests successively the servers specified in the response of the requests until reaching the searched record.
%zone rather than the location identifier  \texttt{k1.example.com.} itself. 
In the example request to look for ``data.k1.example.com'', the client then requests the server responsible for the \text{com.} zone before being able to request the server responsible for the zone \texttt{example.com} knowing that the data location identifier storing the data is \texttt{k1.example.com}.

The approach of DNS leads to better scalability properties because the table of names (addresses) stored in each server is reduced (due to the distributed nature of the processing) compared to a centralized approach.
Nevertheless, this reduction is not homogeneous and depends on the balance of the tree and the balance of the data identifiers used. In the worst case, one centralized server stores on the order of $O(d)$ records whereas, in the case of a balanced tree, each server only stores $O(d/n)$ location records.

On the other hand, this approach is not resilient to physical network topology changes because if the address of a server changes, 
it must be reflected on the server managing the parent zone~\cite{1605858}.
Similarly, if a server in the tree becomes unreachable,
all the records stored in the corresponding subtree cannot be accessed
%Similarly, if a server in the tree becomes unreachable, all the data fragments whose identifiers are stored in the corresponding subtree cannot be accessed anymore.
For instance, in Figure~\ref{fig:dns_resolution}, if the server managing the zone \texttt{com.} at \texttt{addr2} becomes unreachable, all the servers in the coresponding subtree and thus all the records belonging to a subzone of \texttt{com.} cannot be reached.

Adding resiliency to dynamic physical network topology changes cannot be done without adding extra complexity. This is because servers do not necessarily know the entire network topology within a short time. Therefore, servers cannot join their parents to change the address specified for the zone they manage.
For instance in Figure~\ref{fig:dns_resolution}, if the server managing the zone \texttt{example.com} is moved from \texttt{addr4} to a new address, the server managing the zone \texttt{com.} at \texttt{addr2} has to be updated to account for this change in the network. From a limitation point of view, we identify that the suffix needed to aggregate the data identifier within different servers is a drawback because the users cannot really choose the data identifier they want to use.

%to choose the server the location has to be stored.
%In a distributed environment, the server storing the location of the data can be located far away from the data itself.

\subsubsection{Gossip protocols}

\begin{figure}[t!]
 \centering 
 \includegraphics[width=\columnwidth]{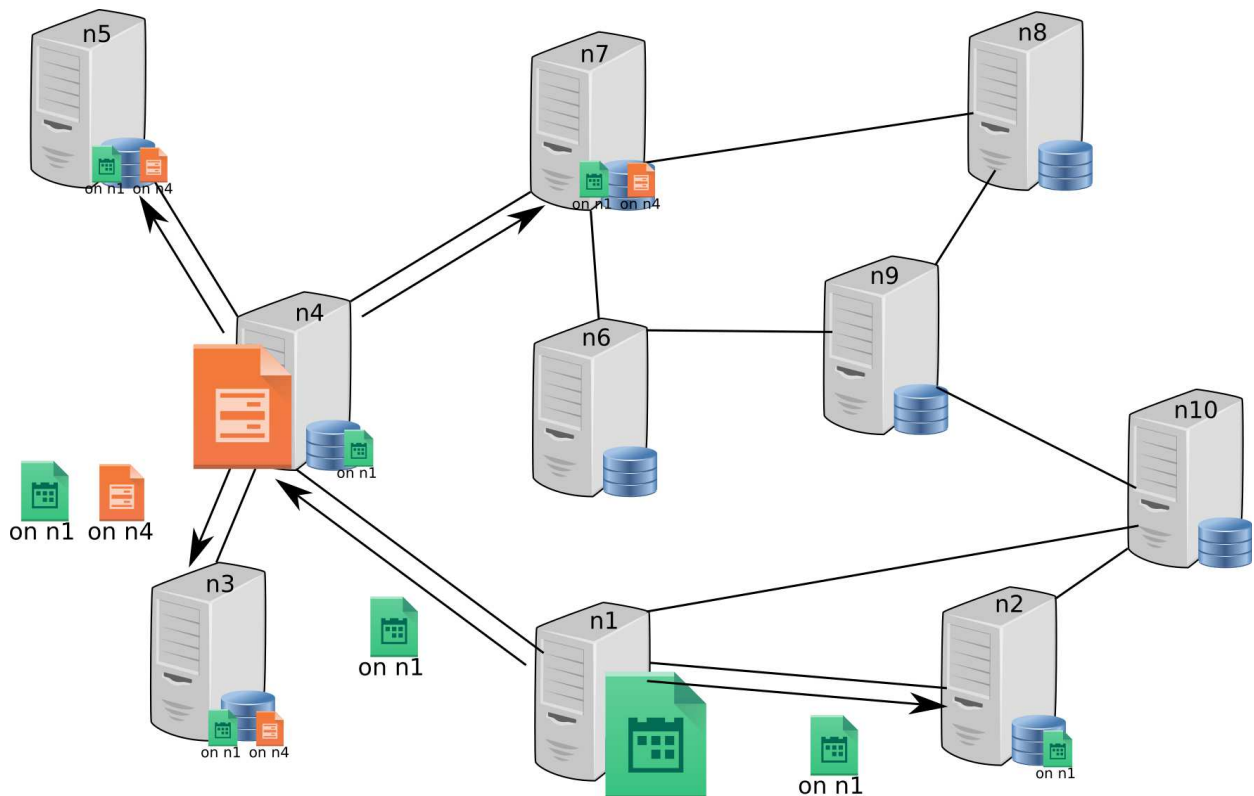}
 \caption{Example of a Gossip protocol where ``n1'' stores a green file and ``n4'' stores an orange file. By propagating the knowledge, all nodes know the location of these two files.}
 \label{fig:gossip}
\end{figure}

An alternative way to overcome the bottleneck of a centralized server is to replicate the database on every node.
This is what the Gossip protocols do in general. On a set of regular intervals, each node contacts its neighbors to exchange the list of identifiers for the data fragments they store. They also exchange the location identifiers learned from other nodes. This way, each node is able to construct its own table to locate every data fragment in the network.
Such a protocol is illustrated in Figure~\ref{fig:gossip}.
The main advantage of this approach is that the location identification process consists of using this local table only without exchanging network traffic and hence the bottleneck.
However, the scalability issue of the centralized approach is not actually solved because when a large amount of data is stored, the table stored by each node also increases proportionally.

%\subsection{Gossip}
Cloud solutions like Dynamo~\cite{10.1145/1323293.1294281} utilizes such an approach. In addition, all of the solutions relying on a blockchain technology~\cite{10.1007/978-3-319-59665-5_15} adapt this approach because in such solutions, the blockchain is replicated on all the nodes and stores the location of each data fragment.

\subsubsection{Names embedding the location}\label{subsec:location_embedded}
Yet another solution is to embed the database inside the 
data identifiers themselves, as it is shown in Figure~\ref{fig:classification}.
For instance, accessing a data named  \textit{http://<ip address>/data} or even email addresses embed the name of the server inside the identifier. In order to reach a URL such as \textit{http://<ip address>/data}, the network will first route the request to the IP address and then, once the server is reached, the ``data'' will be queried.
The main drawback of such an approach is the addition of a constraint on the data naming process. Moving a data fragment over the network leads to a name change while it allows us to easily locate the data. Nevertheless, routing the request to the destination is still a major challenge.

In IP routing schemes, each router reads the entire routing table (though might be sizable) to determine where to forward the request.
The selected route is the most specific one matching the destination address. For instance, if a router has  two destination addresses: \texttt{2001:db8:1::/64 to routerA} and \texttt{2001:db8::/32 to routerB}, a packet destined to \texttt{2001:db8:1::1} would be forwarded to \texttt{routerA} due to a more specific matching with the first option.

In order to limit the size of such tables, two main mechanisms are utilized in IP networks. The first mechanism is the prefix aggregation which merges several routes pointing to the same next hop. For instance, if a router stores the routes \texttt{2001:db8:abcd:1234::/64 via 2001:db8::} and \texttt{2001:db8:abcd:12ab::/64 via 2001:db8::}, the two routes can be aggregated into a more general single route, namely \texttt{2001:db8:abcd:1200::/56 via 2001:db8::}.
This process leads to a reduction in the number of routes stored in each router with the goal of increasing performance and scalability.
Nevertheless, finding the minimal set of routes that do not create loops in the network is an NP-complete problem~\cite{Le:2011:RA:2079296.2079302} and a few heuristics' have already been proposed~\cite{SRIDHARAN2001111}.

The second approach is to use a default route.  Routers only know the routes to few IP prefixes (generally those which are administered by the same party) and forwards all the packets to a default route.
Although this leads to a reduction in the number of routes in the core routers, it does not limit the number of routes in edge routers which still have to store the full routing table. Today, the full routing table in IPv4 stores more than 780\,K routes\footnote{\url{https://bgp.he.net/report/netstats}} which clearly illustrates the limitation of both approaches.

\subsection{Hash-based protocols}

A second way to locate data is to rely on a predefined mathematical function (usually through utilization of a one-way hash function) applied to the  data identifier.

\subsubsection{Consistent Hashing-based approaches}

Consistent Hashing (CH)-based approaches such as CRUSH~\cite{Weil:2006:CCS:1188455.1188582} or SWIFT's RING~\cite{Arnold:2014:OSU:2755277} propose to take into account the placement constraints in the corresponding mathematical function.
Although such approaches provide partial control over the data placement process, they require each network node to be aware of the entire network topology to be able to map a hash value to the node address (size of the routing tables is on the order $O(n)$).
The advantage is that because all nodes know the entire network topology, they just have to hash the data identifier and map it on the table to be able to find which nodes are the destination for the data. The data locating process sends only one request on the network $O(1)$, even if this request has to go across all the network.

Sharing the entire topology with all the network nodes makes it hard for the system to face alterations in the topology while at the same time satisfying minimum network traffic. This is because it would require to redistribute the new topology information among all the nodes ($O(n)$). Finally, we would like to note that the constraints on data placement and naming are the same as in the DHT because these constraints are inherent to the use of hash functions.
These characteristics are summarised in Table~\ref{tab:summary}.

\subsubsection{Distributed Hash Tables}

\begin{figure}[t!]
 \centering 
 \includegraphics[width=\columnwidth]{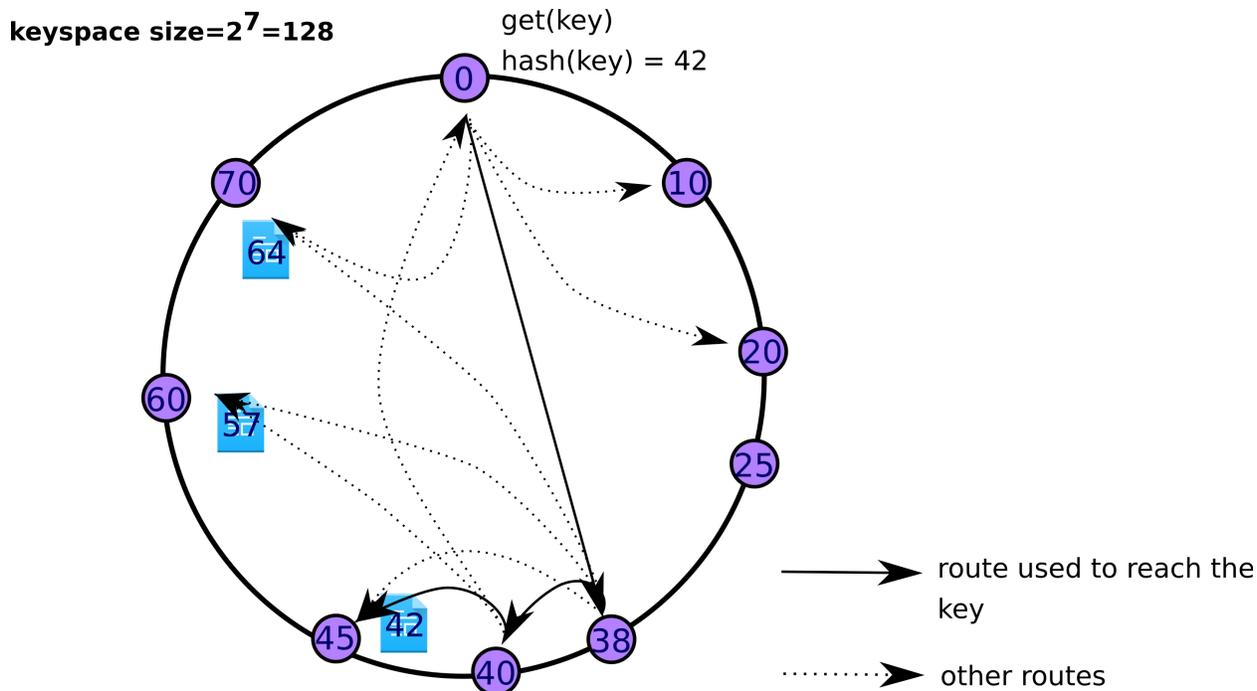}
 \caption{Example of a DHT whereby a hash function is used to determine the key \texttt{key} which is stored on the node with the identifier of 42.} \label{fig:chord_example}
\end{figure}

Distributed Hash Tables (DHT) locate the node storing a specific data fragment by hashing the
name of its identifier (for instance a cryptographic function SHA-1 in the Chord DHT~\cite{Stoica:2001:CSP:964723.383071}). Figure~\ref{fig:chord_example} shows an example of a DHT where the key \texttt{key} is stored on the node with the identifier immediately greater than 42. This location identifier is found by the client before sending any request. Note that the  process of locating a data fragment is completely local as it only consists of computing an appropriate hash value.

\begin{figure}[!b]
 \centering
 \includegraphics[width=\columnwidth]{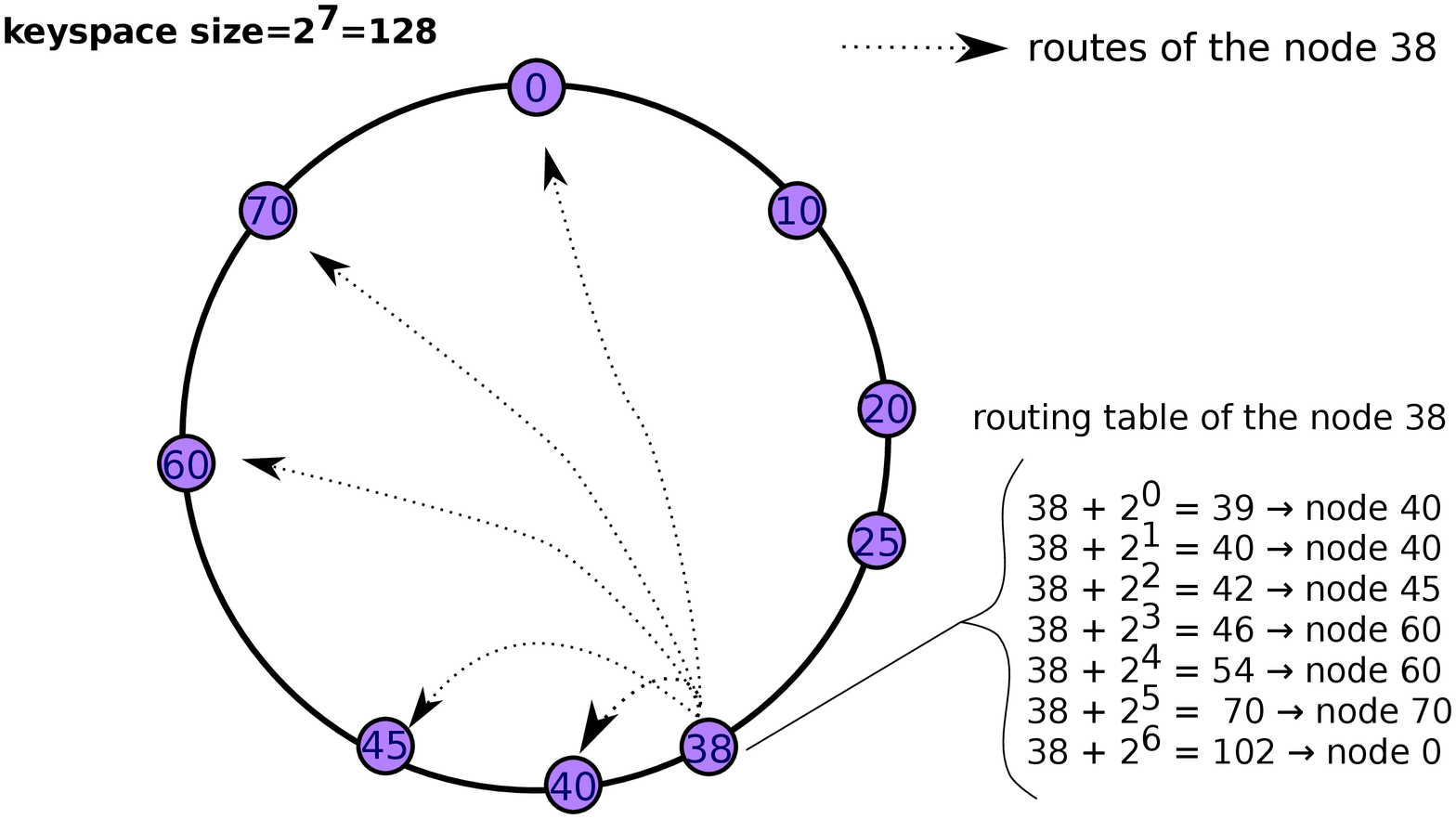}
 \caption{A typical routing table in a DHT.}
 \label{fig:dht_route}
\end{figure}

On the other hand, the DHT possesses a routing mechanism to reach the destination node storing the data. The routing table contains the address of a particular node's identifier and the node forwards the request to the node with the closest identifier to the destination identifier. This way, since identifiers are sorted in the routing table, there is no need to browse all possible routes. In addition, limiting the number of routes helps build a scalable system. In a DHT, each node keeps at most $\log(N)$ routes where $N$ is the size of the keyspace. For instance, in Chord DHT, each route points to the node with the identifier immediately superior to
$node\_identifier + 2^{i}$ where $i$ varies from 0 to $\log(N)$. This is illustrated in  Figure~\ref{fig:dht_route}.

%\textcolor{red}{Chord is a bit old. Are there any improved version and more recent work for Chord or other DHTs in literature?} % Thanks I added a reference to Pastry and Kademlia in the following paragraph

The inherent design of DHT is scalable because each node of the network stores a routing table with at most $O(log(n))$ entries.
Nevertheless, having a small routing table leads to an increase in the number of hops and thus reducing the overall data access latency performance~\cite{10.1007/978-3-540-30183-7_9,4146982}. 
Several DHT implementations have been proposed to reduce the number of hops such as Kademlia~\cite{maymounkov2002kademlia} or Pastry~\cite{Rowstron:2001:PSD:646591.697650}. These solutions increase the number of neighbors in order to reduce the number of hops. In fact, we realize that this performance issue (access latency) can be tied to the Brewer's CAP theorem and its trade-off between availability and consistency. To overcome this performance issue, many storage solutions have preferred to relax the strong consistency requirement. This is one of the reasons why implementations often rely on the eventual consistency model instead~\cite{Lakshman:2010:CDS:1773912.1773922}.

This approach also supports dynamic changes in the network topology as it is able to automatically reconfigure. Adding a node to the network leads to inform the  $O(log(n))$ neighbors and to re-balance the data.
Because of the use of a hash function which is consistent, data balancing leads to move the minimum set of data on the order of $O(d/n)$ where $n$ is the number of nodes and $d$ is the replication level. 

However, the drawback of this approach is that it is not possible to choose the node storing the location record for a specific data fragment. In that sense, it is possible that the location record can be stored far away from where the data is originated.
In other words, in order to store a data fragment on a specific node, you need to find a data identifier that has a hash value that fits within the range of keys stored by that node.
It also means that moving a data fragment would lead to a compulsory change in the data identifier.

\subsection{Exploration-based protocols}
The exploration-based protocols are based on clients randomly exploring the network and trying to find the data they are looking for. The main advantage of this approach is that it does not require the storage of the data location. There is no stored record to indicate that a specific data fragment is stored on a specific node just like we typically have in a local database approach. 
In addition, exploration methods do not impose any constraints on the location, unlike computation-based strategies.

\subsubsection{Flooding}
\begin{figure}[!t]
 \centering 
 \includegraphics[width=\columnwidth]{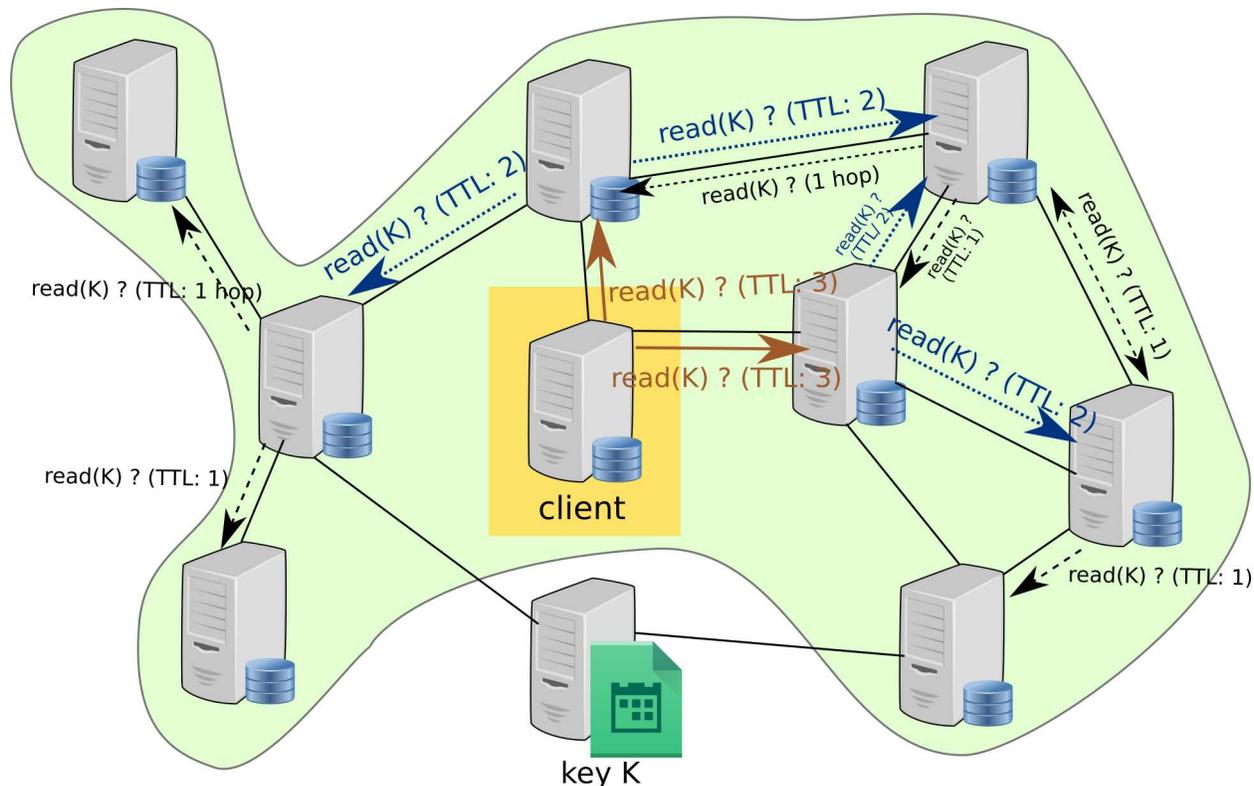}
 \caption{Example of a flooding approach when the Time-To-Live (TTL) specified is too low to reach the needed data object.}
 \label{fig:flooding}
\end{figure}

In flooding approaches, when a user looks for a piece of data fragment, nodes of the network request their neighbors whether they store the data fragment the user is asking for.
If not, nodes forward the request to their neighbors. 
The process continues until the requested data fragment has been found or until all nodes have received the request. A Time-To-Live (TTL) associated with the request limits the loops.
An exploration method based on flooding is illustrated in Figure~\ref{fig:flooding}. The biggest advantage of this approach is the absence of maintenance for forwarding tables. Each node is aware of its neighbors only where network connections are available. 

\subsubsection{Random walk}

Random walk approaches appeared in literature as an improvement to flooding approaches~\cite{1354487}.
In a random walk scheme, instead of forwarding a request to all its neighbors, each node chooses one or several specific neighbors which are the most promising.
The ultimate goal is to locate the data easily on time without overloading the network links.

To select the relevant neighbors, several algorithms and approaches have been proposed. Some of them rely on bloom filtering~\cite{8264842} while others use complex heuristics~\cite{10.1145/1921206.1921230}.
Random walks strategies are used in storage solutions like GNUTella~\cite{990433}. In practice, this approach relieves the scalability issue of the flooding approach but in the worst case, it potentially requires to contact almost all the network nodes. This approach is used in different software implementations such as GNUTella ~\cite{bordignon2001gnutella} or Spinneret~\cite{4228383}.

%\textcolor{red}{(Are there any more exploration-based approaches? Randomwalk is listed as the first? - Bastien: I separated flooding from random walk as it is done in some papers.)}

\subsection{Discussion about the existence of other approaches}

In this article, we focused on the aforementioned three mechanisms used to locate data, because we keenly believe that all the existing approaches in literature either fall under one of them or are simply a mixture of two or more mechanisms.
%To the best of our knowledge, these presented mechanisms are the only existing methods.
%This we believe is supported by the fact that some protocols do not even define a specific implementation.
For instance Cassandra\cite{10.1145/1773912.1773922} relies on a one-hop DHT
which is in fact a DHT where data is placed according to a hashing algorithm
and where node identifiers are distributed with a gossip mechanism.
Such a solution is intended to achieve a tradeoff between the benefits and the drawbacks of the two mechanisms involved.
%Again, to our knowledge and analysis, all such implementations seem to rely on the three mechanisms previously presented. 
Let us detail two specific examples of a mixture of mechanisms:

\subsubsection{Information Centric Network}
 Information Centric Network (ICN) is a new approach that is deemed to replace an IP network. In the ICN, users do not contact a specific node but instead, they send their requests for a specific data fragment they would like to retrieve. It is the network that is in charge of routing the request to the intended node.
 
 Many implementations have been proposed and they all rely on a local database. For instance, in Data-Oriented Network Architecture~\cite{Koponen:2007:DNA:1282427.1282402} (DONA), routers store the association between a data identifier and the next hop in the network. This approach clearly relies on a local database concept that has the same scalability issues mentioned earlier. Similarly, Mobility First~\cite{Seskar:2011:MFI:2089016.2089017} proposes to store all the routes in a central server whereas
 COntent Mediator architecture for content-aware nETworks (COMET)~\cite{salguero2010content} and Naming Data Network (NDN) propose to store the routes in a hierarchical, DNS-like infrastructure.

\subsubsection{Location Identifier Separation Protocol}

The second protocol we shall focus on is Locator Identifier Separation Protocol (LISP). Standard IP protocol uses IP addresses that represent not only the location of a node (the set of networks it is connected to) but also its identifier. This is exactly what we described in Section~\ref{subsec:location_embedded}.
 
The LISP protocol separates these two roles by introducing an association between a node identifier and its location. A variety of implementations were proposed. For instance, LISP-DHT~\cite{10.1145/1544012.1544073} relies on a DHT associated with a computation function. Another implementation called LISP-Tree~\cite{5586446} relies on a local database and more specifically on a DNS-like protocol. We also note that the implementation of Saucez~\textit{et al.}~\cite{6279445} is inspired by the Open Shortest Path First protocol (OSPF) which is quite similar to a gossip protocol in the sense that the entire database is replicated on all nodes.

Through these two examples, we demonstrated that the three proposed approaches are quite commonly used in different fields of networking and distributed computing. We realize that these two protocols do not have an implementation that uses an alternative way to locate data or to make the association between identifiers.

\section{A proof of the conjecture}\label{sec:proof}

In this section, we formulate the conjecture that no distributed storage system can be scalable, tolerant to network partitioning and letting the users choose the name of data they store without constraints. We then introduce existing fundamental distributed system trade-offs that we shall rely on, before finally establishing the proof of our conjecture in a separate subsection.

\subsection{Formulation of a fundamental conjecture}

We have summarized the advantages and drawbacks of different solutions in Table~\ref{tab:summary} in terms of $O(.)$ notation for the three classes of concepts we identified and presented previously.
% Try to explain the table with more details
%\textcolor{red}{Bastien: More explanations about the table:}
Directory-based approaches like using a central server or a gossip protocol
face a limitation in the size of the routing table, which grows with the number of objects stored ($O(d)$).
However, in case of a request, it is easy to locate a data fragment as all the information is centralized. It takes $O(1)$ network requests to reach the central server and $O(0)$ requests to a gossiped database because in a gossip protocol, the database was previously replicated on all nodes to enable local access.
We also note, that the central server may be located at the other side of the network and therefore the only request needs to be forwarded by $O(diameter(n))$ links (where $n$ is the number of nodes). 
These approaches also allow us to add more nodes easily as the nodes just need to know the address of the central server or a few neighbors to start gossip.

Approaches like DNS or IP routing try to face the scalability limitation by introducing a hierarchical approach. This hierarchical approach comes with a hierarchical naming system and thus constraints on naming.
The only difference in DNS and in IP routing is that DNS is an iterative protocol, and IP routing is a recursive protocol. Therefore, the number of requests and the number of network links used vary.

In Hash-Based protocols, the constraint on naming and placement is very strong.
In a DHT, it requires $O(log(n))$ requests to locate a piece of data. Because the overlay network does not map the physical topology, each request can lead to cross the entire network. Thus, the number of network links used to be $O(log(n)$ $\times$ $diameter(n))$.
In  Consistent Hashing, each node needs to know all of the nodes composing the network, thus a routing table of $O(n)$ size is required.

Finally, in exploration-based protocols, there is no database but it requires in the worst case that each of the $n$ nodes contacts each of its neighbors. In the worst case, the network is a complete graph, thus $O(n^2)$ links are required to be used.
%end

Table~\ref{tab:summary} clearly demonstrates that 
none of the presented solutions is able to (1) scale with the number of data objects stored (there is no single point in the network that needs to store as many records as data objects in the network), (2) possesses a form of tolerance to the physical network topology changes and (3) enable users to choose freely where to place their data and what the data identifier would be while storing it\footnote{Note that we do not strictly mean physical location here. In other words, the user might be interested in the properties of the location s/he chooses such as privacy, durability, access privileges etc. Especially in cloud settings, the user may require such properties to hold for the data and may wish to assign a specific name for the data for compatibility reasons. Here the user can be thought as the cloud admin as well, it does not need to be a third party or a customer receiving service. As the name of the file may change over the course of its lifetime as stored in the system. However, the question is about whether or not there would be any connection between the naming convention and the storage location. When we refer to the user having all freedom to choose a name, we strictly mean that this link is broken.}.
In other words, it shows that none of the solutions is able to satisfy the three properties we have presented in Section~\ref{sec:properties} all at the same time.

Based on this observation, we conjecture that it is impossible to design a distributed data storage solution that can simultaneously meet these three properties mentioned above.
This claim is similar (but we believe orthogonal) to the popular Brewer's CAP theorem, which establishes that it is not possible to have a sequential consistency, availability, and support for network partitioning all at the same time~\cite{Gilbert:2002:BCF:564585.564601}.

\begin{table*}

\begin{center}
\vspace{2cm}
  \begin{adjustbox}{addcode={\begin{minipage}{\width}}{\caption{%
     Table showing how existing solutions deal with scalability, tolerance to physical network topology changes and constraints on naming or placement of the data for a network of $n$ nodes.
      }\label{tab:summary}\end{minipage}},rotate=90,center}
\begin{tabular}{|ll|l|l|l|l|l|l|l|}
\rot{~} & \rot{~} & \rot{Size of a routing table} & \rot{Number of routing tables} & \rot{\makecell{Number of requests\\to locate a piece of data}} & \rot{\makecell{Number of network links used\\in the data location process}} &
\rot{\makecell{Network traffic to\\add or remove a node}}  &
\rot{\makecell{How to name a piece of data\\that we want to be placed\\on a server ``x''? }} &
\rot{\makecell{Data identifier changes\\when the data is moved}}  \\
\hline
\multicolumn{2}{|c|}{Solution} & \multicolumn{4}{c|}{\cellcolor{y_color}Scalability} & \multicolumn{1}{c|}{\cellcolor{g_color} \Gape[0pt][0pt]{\makecell{Tolerance to\\Network Topology\\Changes}}} & \multicolumn{2}{c|}{\cellcolor{b_color} \Gape[0pt][0pt]{\makecell{Constraints on Naming\\or Data Placement\\}}} \\
\hline
\multicolumn{2}{|l|}{\textbf{Directory-based protocols}}  & \cellcolor{y_color} & \cellcolor{y_color} & \cellcolor{y_color} & \cellcolor{y_color} & \cellcolor{g_color} & \cellcolor{b_color} & \cellcolor{b_color} \\  

~ & Central server & \cellcolor{r_color}$O(d)$ & \cellcolor{y_color}$O(1)$ & \cellcolor{y_color}$O(1)$ & \cellcolor{y_color} \Gape[0pt][0pt]{\makecell{$O(diameter(n)$ \\$=O(n)$}} & \cellcolor{g_color} $O(1)$ & \cellcolor{b_color} ``obj'' & \cellcolor{b_color} no \\

~ & Gossip & \cellcolor{r_color}$O(d)$ & \cellcolor{y_color}$O(n)$ & \cellcolor{y_color}$O(1)$ & \cellcolor{y_color}$O(0)$ &
\cellcolor{g_color}$O(0)$ & 
\cellcolor{b_color} ``obj'' & \cellcolor{b_color} no \\

~ & Domain Name System & \cellcolor{y_color} \Gape[0pt][0pt]{\makecell{$[O(d/n);$\\$O(d)]$}} & \cellcolor{y_color}$O(n)$ & \cellcolor{y_color} $O(diameter(n))$ & \cellcolor{y_color} \Gape[0pt][0pt]{\makecell{$O(2 \times$\\
$diameter(n)^{2})$}} &
\cellcolor{r_color} \Gape[0pt][0pt]{\makecell{does not support\\ dynamicity}} & 
\cellcolor{o_color} ``obj.suffix'' & \cellcolor{b_color}no \\

~ & IP Routing & \cellcolor{y_color}$O(n)$ & \cellcolor{y_color}$O(n)$ & \cellcolor{y_color}$O(1)$ & \cellcolor{y_color}$O(diameter(n))$ &
\cellcolor{r_color} \Gape[0pt][0pt]{\makecell{an additional\\protocol is required}} &
\cellcolor{o_color} ``destination/obj'' & \cellcolor{r_color} yes \\

\multicolumn{2}{|l|}{\textbf{Hash-based protocols}} & \cellcolor{y_color} & \cellcolor{y_color} & \cellcolor{y_color} & \cellcolor{y_color} & \cellcolor{g_color} & \cellcolor{b_color} & \cellcolor{b_color} \\  
~ & Distributed Hash Tables & \cellcolor{y_color}$O(d/n)$ & \cellcolor{y_color}$O(n)$ & \cellcolor{y_color}$O(log(n))$ &  \cellcolor{y_color} \Gape[0pt][0pt]{\makecell{$O(log(n) \times$\\$ diameter(n))$ \\$=O(n \times log(n))$}} & %& \cellcolor{b_color}$hash^{-1}(obj)$ & yes \\
\cellcolor{g_color}$O(log(n)+d/n)$ &
\cellcolor{r_color}$hash^{-1}(obj)$ & \cellcolor{r_color}yes \\

~ & Consistent hashing & \cellcolor{y_color}$O(n)$  & \cellcolor{y_color}$O(n)$ & \cellcolor{y_color}$O(1)$ & \cellcolor{y_color} \Gape[0pt][0pt]{\makecell{$O(diameter(n))$\\$=O(n)$}} &
\cellcolor{g_color}$O(n+d/n)$ &
\cellcolor{r_color}$hash^{-1}(obj)$ & \cellcolor{r_color}yes \\

\multicolumn{2}{|l|}{\textbf{Exploration-based protocols}} & \cellcolor{y_color} & \cellcolor{y_color} & \cellcolor{y_color} & \cellcolor{y_color} & \cellcolor{g_color} & \cellcolor{b_color} & \cellcolor{b_color} \\
~ & Flooding & \cellcolor{y_color} $O(0)$ & \cellcolor{y_color} $O(0)$ & \cellcolor{r_color}$O(n)$ & \cellcolor{r_color} \Gape[0pt][0pt]{\makecell{$O(n \times n)$\\$=O(n^2)$}} & 
\cellcolor{g_color} $O(0)$ &
\cellcolor{b_color} ``obj'' & \cellcolor{b_color} no \\

~ & Random walk & \cellcolor{y_color} $O(0)$ & \cellcolor{y_color} $O(0)$ & \cellcolor{r_color}$O(n)$ & \cellcolor{r_color} \Gape[0pt][0pt]{\makecell{$O(n \times n)$\\$=O(n^2)$}}  & 
\cellcolor{g_color} $O(0)$ &
\cellcolor{b_color} ``obj'' & \cellcolor{b_color} no \\

\hline

\end{tabular}
\end{adjustbox}
%\caption{Table showing how existing solutions deal with scalability, tolerance to physical} network topology changes and constraints on naming or placement of the data for a network of $n$ nodes.} %\textcolor{red}{This table may need more explanation.}}
\end{center}
\end{table*}

\subsection{Background: Existing fundamental trade-offs for distributed systems}

\begin{figure}
\begin{center}
  \subfloat[\label{subfig:cap}][-- CAP Theorem]{
   \includegraphics[width=.48\columnwidth]{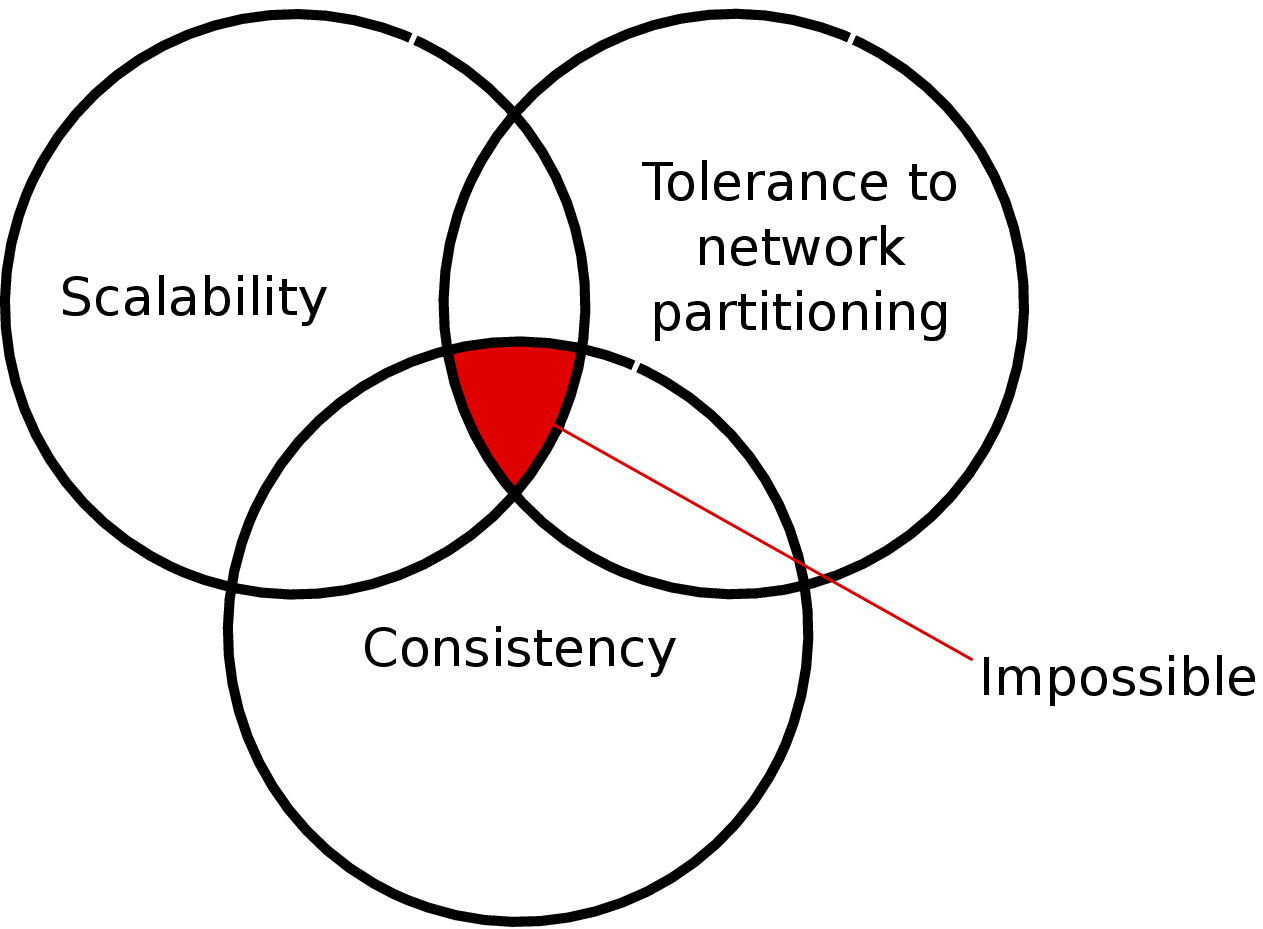}
   \label{subfig:cap}
  }
  \subfloat[\label{subfig:flp}][-- FLP Theorem]{
   \includegraphics[width=.48\columnwidth]{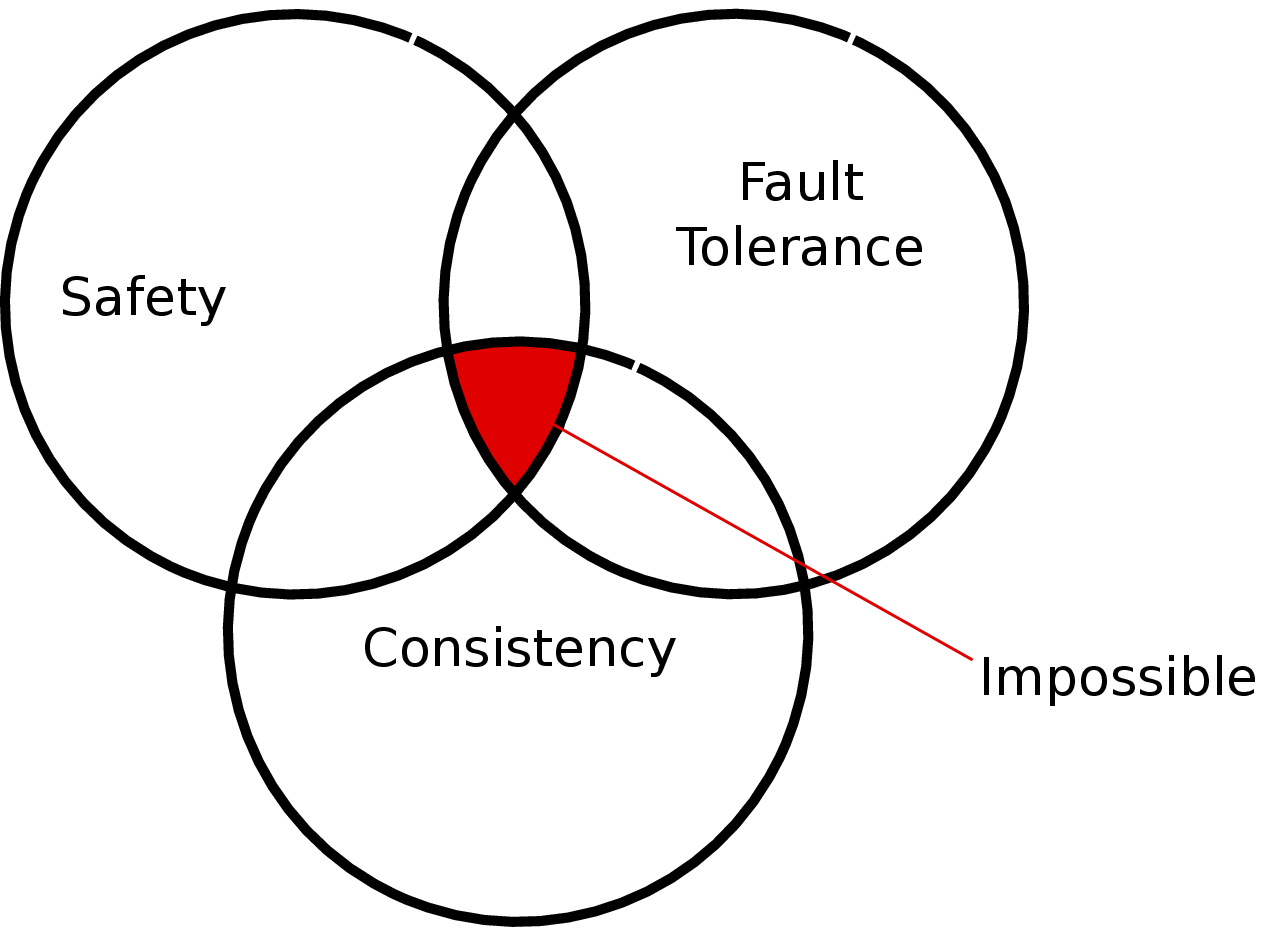}
   \label{subfig:flp}
  }
 \caption{An illustration of the impossibility results of the famous CAP (a) and the FLP (b) theorems.}
 \label{fig:tradeoff}
\end{center}
\end{figure}

The conjecture we had in the previous section is not the first fundamental trade-off in distributed systems sharing these three properties.
As shown in Figure~\ref{fig:tradeoff}, there exists known fundamentals trade-offs such as the Fisher Lynch and Paterson~\cite{10.1145/3149.214121} trade-off or the well-known Brewer's theorem~\cite{Brewer:2010:CFT:1835698.1835701}.
These trade-offs state that
any distributed system can at most satisfy two out of the three important distributed system features: Consistency, Availability, and Partition Tolerance. However, other important properties need to be considered together in a distributed system design. 

\subsubsection{Property definitions}

In addition to the properties we defined in Section~\ref{sec:properties}, we need to formally define some more additional properties for convenience.

 \paragraph{\textbf{Consistency (C)} }
In our context, we assume consistency to mean strong consistency (linearizability) just like in CAP. This means that all operations can be linearized in total ordering.
In other words, any read operation that begins after a write operation completes must return that value, or the result of a later write operation.

\paragraph{\textbf{Safety}}
The safety is a relaxation of consistency.
In a safe system, nodes have to agree on a consensus of the value returned.
But, contrary to Consistency (C), Safety does not guarantee the value returned to be the last written one.

%\textcolor{red}{(I think consistency and safety are not referring to the same thing.)}

\paragraph{\textbf{Availability  (A)} and \textbf{Liveness}}
An available system is a system in which any request must eventually be addressed. 
However this notion of availability
referred in the Brewer's theorem~\cite{Gilbert:2002:BCF:564585.564601} is not very clear because it is not specified what amount of delay is tolerable.
For instance, it may be possible to wait for an unlimited amount of time, turning the system not live.
%In the following of the proof, the availability is assumed to be the same as liveness.
In order to clarify this, we define in the following the $\tau$-availability ($\tau$-A) to characterize systems with a response given within a
threshold time $\tau$.
We note that if $\tau \to \infty$ we obtain the liveness property which would be a special case of this availability definition.

 \paragraph{\textbf{Partition tolerance (P)}}
The network partition tolerance refers to the fact that due to unexpected node failures, the network can  arbitrarily be
split into several connected components called network partitions and messages sent between partitions are permanently lost.

\paragraph{\textbf{Fault tolerance (F)}}
The fault tolerance is the ability of a system to continue its normal operation without interruptions when a subset of nodes fails. This can be implemented in the form of a hardware or a software fault mitigation system.

\subsubsection{Fisher, Lynch and Paterson (FLP) theorem}

One of the fundamental trade-offs in distributed systems is the impossibility theory of Fisher Lynch and Paterson (FLP)~\cite{10.1145/3149.214121}.
This theorem states that in a reliable network, it is impossible to have a deterministic algorithm to solve the consistency problem in
case of any node failure or in case of an asynchronous system model. In other words, no distributed system can possess safety, liveness, and fault tolerance properties all at the same time.

\subsubsection{Brewer's CAP theorem}

A second fundamental trade-off is the Brewer's CAP theorem~\cite{Brewer:2010:CFT:1835698.1835701}.
This theorem states that in an asynchronous network model, it is not possible to create a distributed solution
that is tolerant to network partitioning, available, and sequentıally consistent, all at the same time.

The consequence of the CAP theorem is that all distributed data storage solutions have to make a choice.
Either they guarantee strong consistency, they enable users to
access the latest version of the data without guaranteeing a quick response in case of a network partition (CP) 
or they stay available (quick response) when some of the nodes fail without guaranteeing a strong consistency (AP). Under the light of this observation, we will state our theorem by considering these two kinds of systems separately.

While this theorem is criticised~\cite{6122007,Huang2012}, especially for its lack of definitions for the properties, it is still a viable and a fundamental reference to characterise distributed systems.
%The drawback of this theorem is that the terms used are not clear and sometimes not defined.
%It required us some 

\subsection{A Proof}
\begin{figure}
\begin{center}
 \includegraphics[width=.7\columnwidth]{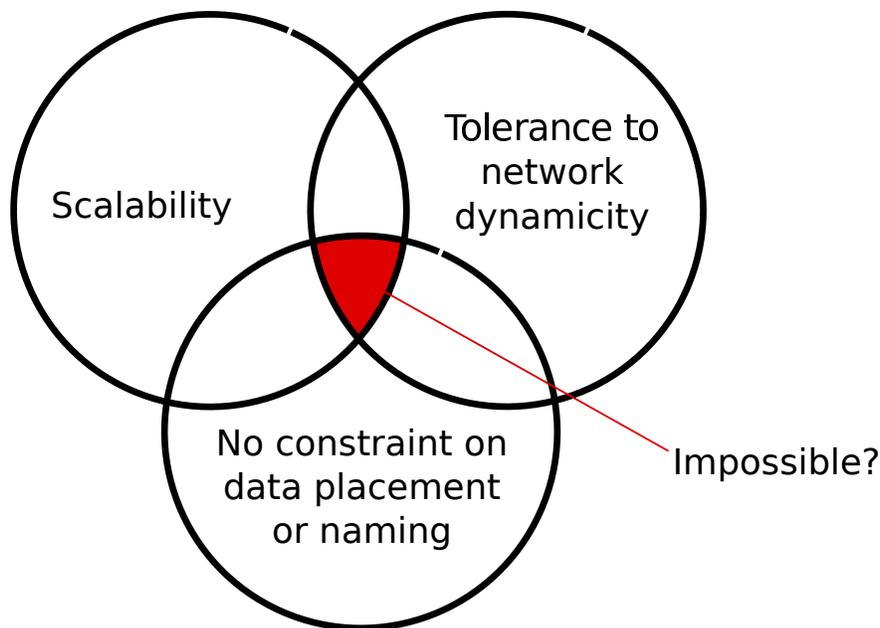}
 \caption{An illustration of the impossibility result we desire to prove.}
 \label{fig:s-To-N}
\end{center}
\end{figure}

In this subsection, we will prove that no distributed data storage solution can meet simultaneously all three
properties i) Scalability (S), ii) Tolerance to Dynamicity of the Network topology (To) and iii) the absence of constraints on data placement/naming i.e., No Constraint on data placement/Naming (N). This is illustrated in Figure~\ref{fig:s-To-N}.

%\paragraph{\textbf{Scalability (S)}}

To prove this conjecture, we primarily take in to account the implications of the CAP theorem. In other words, since partition tolerance is an inevitable requirement of any distributed system, we consider two potential cases.
In the first case, we prove the impossibility result for Available (A) systems that can also deal with Network Partitioning (P). 
In the second case, we will establish the proof for the systems that possess Consistency (C) and the tolerance to Network Partitioning (P) properties at the same time.

We emphasize that we do not consider solutions that are Available (A) and strongly Consistent (C) without any tolerance for Network Partitioning (P) because this last property is essential in distributed systems as soon as the storage solution is spread among several geographical locations. Another argument for not considering solutions that are available and consistent has been exposed in an extension of the CAP theorem.
Abadi~\textit{et al.}~\cite{10.1145/3197406.3197420} considers that network partitioning is not a characteristic of the system but a characteristic of the environment. The system's property is its reaction to this partitioning. They propose the PACELC model that can be summarised in the following way:
When the network is partitioned (P), there is a trade-off between Availability (A) and Consistency (C). But, when the network forms a connected component, %(Else - E), 
the trade-off is between Latency (L) and Consistency (C).
Finally, Diack~\textit{et al.} showed that it is impossible to have Consistency and Availability simultaneously~\cite{Diack_Ndiaye_Slimani_2013}.
For these three reasons, we do not consider a non-practical solution that would be both available and consistent.

\subsubsection{\textbf{A Proof for Available and Network Partition Tolerant (AP) systems}}

In this case, we consider all the possible mechanisms to locate data in a $\tau$-AP distributed system and state a hypothesis.
More specifically, we assume that a system possesses all three features: S, To, and the N. 
In the following paragraphs, we consider different categories of protocols and we show that 
for each one of them, the hypothesis always leads to a contradiction.

\paragraph{Case 1: A Solution relying on a location database}
%Using a local database that stores the location of each piece of data is the first approach. 
%This approach has the advantage that there is no constraint on the names used for the pieces of data (N) and that adding nodes in the network does not impact the existing nodes to access the database (To)
%
%Nevertheless, the scalability is a huge problem of this approach.
%It can be implemented in a centralised way with a central server, leading to having one node receiving all the location requests, creating a Single Point of Failure and a scalability problem.
%It can also be implemented by replicating the database on all the nodes (such as in Amazon Dynamo~\cite{10.1145/1323293.1294281}). The main problem of this is that the creation of a new piece of data
%leads to update the local database of all nodes. Each node has to have the sufficient storage space to store a location record for each piece of data stored in the network.
%This is also a scalability issue.
%
%Therefore, solutions 
%We first suppose a system meets the following characteristics:  $\tau$-AP, Scalable (S), Tolerant To network changes (To) and has no constraints on the data placement or naming (N).

Under the circumstances of the existence of a location database, a node needs to request the database to determine the location of data.
Because the system is both $\tau$-AP and (N), this request is performed at the time a user likes to access a specific data fragment.

However, there exist different ways of setting up the database. Using a centralized database leads all the network flows to concentrate on a specific machine resulting in a system bottleneck. On the other hand, replicating the database on a few nodes can easily be implemented (since strong consistency is not a must) and can solve the aforementioned bottleneck. In both scenarios, however, the database has to store the location of every data fragment which, according to the definition presented above, would violate the Scalability property (S). In addition, the address of the central database should be communicated to all the other nodes for data transfer. So changing the address of the database may be quite difficult without violating the tolerance to the dynamicity of the network topology property i.e., To.

Finally, hierarchical databases, like we have seen in DNS where a ``tree'' approach is adapted can be regarded as a potential solution. In that, a central node may be first contacted but this time instead of responding with the location of a specific data fragment, it would respond with the address of the next database to contact to.
Nevertheless, the root node should still be contacted at each request and known to all the network nodes.
Such a tree approach may alleviate the bottleneck but does not completely remove it.

%But it cannot store all location information in its local database.
%Likewise there is no master in which all location information is stored for other
%nodes to contact. If it does, it would violate the scalability (S).
%Replicating the database on all the nodes also violates the scalability feature (S) because it requires all the nodes to store a location record for each piece of data which is also not scalable.
%This leads to a contradiction and therefore a $\tau$-AP-S-To-N solution cannot exist if it relies on a local database.
%Also, since system is not necessarily CP, it may end up with the wrong location information due to (To) property. If it is always the right location, then it would violate (To) as well.

\paragraph{Case 2: A Solution relying on a computational method}

For a $\tau$-AP solution, nodes may use a computational method to determine the location of a data fragment. Typically, these computational methods serve as functions and take the name of the data fragment as input and map it to the location identifier of the node storing a data replica. In this particular case, however, the location of each data fragment is completely dependent on its identifier and therefore the computational methods violate the (N) property.

 Note that due to the dynamicity of the network, nodes randomly enter or leave the network. Therefore, a mechanism is required to update the mapping function accordingly and enable the solution to be tolerant to physical network topology changes.
Such mechanisms should rely on a local database or an exploration of the network, and therefore will have the same line of limitations similar to the ones presented in this paper.

Finally, it is interesting to note that some computational methods allow users to define ``constraints'' on the data placement process, allowing them to have a computational method that satisfies the (N) property.
For instance, in CRUSH~\cite{Weil:2006:CCS:1188455.1188582}, the placement rules enable users to choose the node a specific piece of data is stored on.
However, rules are replicated among all the nodes.
Therefore, if the placement of all data fragments is constrained, CRUSH algorithm would not act like a hashing mechanism but more like a local database.

\paragraph{Case 3: A Solution relying on an exploration of the network}

The third approach to finding the location of the data fragment is the exploration of the network. In this case, a node can locate the data fragment location (destination) in a number of hops as long as the Time-To-Live (TTL)  allows.

%\textcolor{red}{(is this defined before? If so, let us remind the reviewer.)}

Let us first assume a system that possesses the following characteristics:  $\tau$-AP, Scalable (S), Tolerant To network changes (To) and has no constraints on the data placement or naming (N).
Since the solution is $\tau$-AP, a finite TTL must be used to guarantee a response in a finite time and hence the availability of the whole system.
But at the same time, there is no guarantee to find the right data fragment location within the specified TTL, making the solution violating the AP characteristics.
Therefore, this argument clearly illustrates the contradiction and a $\tau$-AP-S-To-N solution cannot exist when the underlying protocol relies on the exploration of the network.

\subsubsection{\textbf{A Proof for Consistent and Network Partition Tolerant (CP) systems}}

In the previous subsection, we showed that meeting the properties S, To and N all at the same time is impossible for a partition-tolerant distributed data storage system which chooses availability over consistency.
In this subsection, we argue that exact same conclusion can be made for a distributed data storage solution that overweights strong consistency over availability.

%In this second part, we will try to prove that having the S-To-N properties all at the same time is impossible for solutions providing strong consistency.

\paragraph{Case 1: A solution relying on a location database}
Using a similar rationale of the previous subsection, let us assume a system that meets the following characteristics:  Strong consistency (C), Scalability (S), Tolerance to network changes (To) and has no constraints on the data placement or naming (N).
To achieve a strong consistency, there are two options: A local centralized database may be used or it would be required to get the location of all replicas and execute a consensus protocol.
In the first option, the centralized server cannot scale and shall violate the scalability (S) property. In the second option, if multiple nodes have to reach a consensus, the properties S and To both would be required at the same time.
For instance, the Paxos protocol~\cite{DBLP:conf/opodis/Lamport02} requires to have a stable network in a partition that contains more than half of the nodes.
Therefore, it is not possible to achieve the tolerance to network dynamicity (To) property.
To the best of our knowledge, no consensus protocol is able to scale in a dynamic environment. Protocols such as  Practical Byzantine Fault Tolerance (PBFT)~\cite{10.1145/571637.571640} have the same characteristics.

In blockchain architectures, the chain is replicated in all network nodes and each node keeps track of all the old records (blocks) due to the immutability guarantees. Although this leads to serious scalability issues~\cite{8038517}, the process to achieve a consensus when data is added has also its own limitations.
In blockchains, a transaction needs to be checked by several different nodes before its insertion into the chain. Even if the chain itself is not replicated in a consistent manner, the process to insert a new piece of data guarantees consistency. The number of nodes required to validate the transaction creates a trade-off between the performance and the security of the chain~\cite{8962150} (\ie the consistency).
Several solutions have been proposed to overcome these limitations~\cite{8962150,10.1007/978-3-319-39028-4_9}. All these solutions propose to relax the consensus of the read value to maintain acceptable performance and therefore the system would have to violate the consistency (C) property.

%\textcolor{red}{(It may be advisable to talk about non-voting-based protocols as well such as blockchain and Proof of Work Consensus (with few appropriate references) - where still the scalability is an issue.)}.

As a consequence, a system that uses a location database is not plausible if we are to require S, To and N all meet at the same time.

\paragraph{Case 2: A solution relying on a computational method}

In the case of a CP solution, nodes use a computational method to determine the location of a data fragment. Similarly, these computational methods serve as functions and take the name of the data fragment as input and return the location. Since the input is the name of the data, such methods obviously would violate the N property. In other words, the location of a data fragment is completely dependent on its assigned identifier. Similar to AP solution, these mapping functions require each node to know the list of all node identifiers in order to set the range of values the function has to return.

%Rephrasing
Finally, the function needs to be updated each time a node enters or leaves the network. This implies the nodes to be able to determine a consensus about the function to use. Therefore, the function (or a parameter used by the function) can be distributed using a mechanism relying on a local database or an exploration of the network. These mechanisms suffer from the same limitations as the ones presented in this paper.
%End

%\textcolor{blue}{Because of the dynamicity of the network, nodes randomly enters or leave the network.
%Therefore, a mechanism is required to update the function accordingly and to enable the solution to be to network topology changes. However,  such a mechanism should rely on a local database or an exploration of the network, and have the same limitations as the ones presented in this paper.} \textcolor{red}{(This paragraph is identical to the one given for the previous subsection, AP case. Even if we would like to argue the same way, it is better that we rephrase it. Copy-Paste is not a good practice.)}

\paragraph{Case 3: A solution relying on an exploration of the network}

Finally, in the case of an exploration of the network, guaranteeing the strong consistency can be done at either the write or read levels. In both cases, it requires to retrieve or to write many data replicas.
Depending on how replication is performed, achieving consistency may require to retrieve (or write) at least $\frac{r+1}{2}$ times, where $r$ is the number of replicas.
Note that this would imply in the worst case to potentially reach all the nodes in the network, this approach cannot be used with a TTL lower than the diameter of the network. More specifically, we would need to reach at least $\frac{r}{2}$ nodes as required by consensus protocols which would typically lead to contact more nodes because not all network nodes store a data replica. This obviously leads to a scalability (S) issue as the network grows. Also, the use of the consensus algorithm implies that, after a sufficient amount of node failures,
no consensus protocol would be able to operate accurately leading to a failure to simultaneously maintain consistency (C) and the required tolerance to physical network topology changes (To).

\begin{figure}[!t]
 \centering 
 \includegraphics[width=.7\columnwidth]{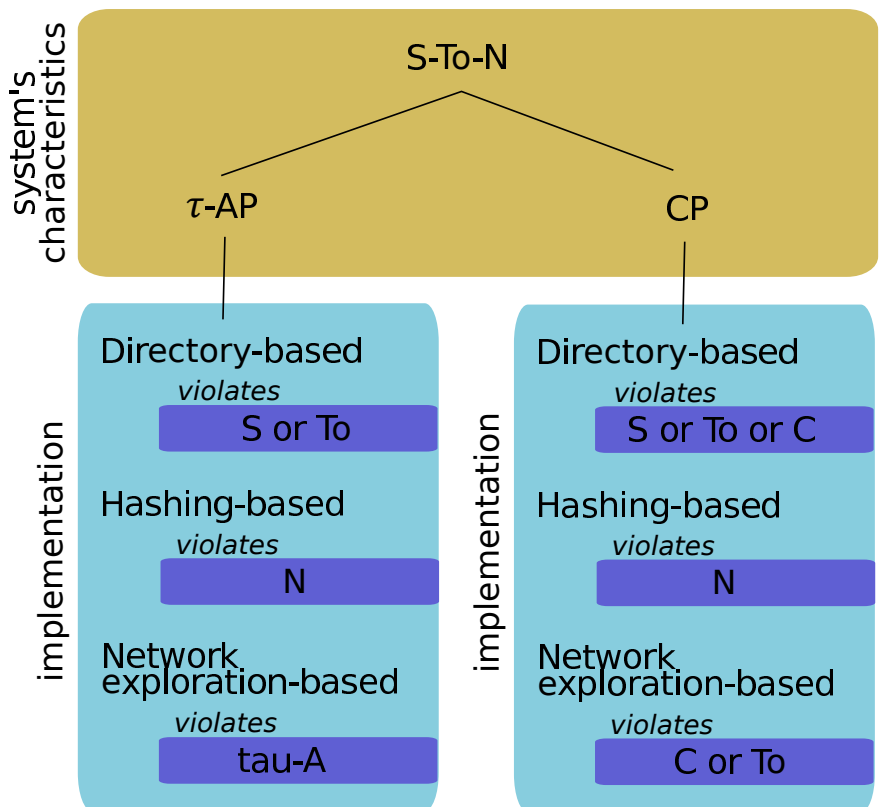}
 \caption{A summary of the proof depending on the system's characteristics considered and the implementation.}
 \label{fig:proof_summary}
 \vspace{-.5cm}
\end{figure}

%P seems to be an inevitable feature of the distributed
%systems otherwise it would not be truly distributed in the right sense. Thus, systems can be AP, CP
%or somewhere in between. In order to achieve for example a ? -AP system we should give up on
%the linearizability and bear with weak consistency models such as eventual consistency. Similarly,
%for any given CP system, it should eventually be available for this system to be of any use. So let
%us call such systems to be useful distributed systems from now on. We conjecture that CAP and
%the following S-To-N trade-off is orthogonal. We state the main theorem as follows.

In conclusion, all of these demonstrations are summarised in Figure~\ref{fig:proof_summary}.
We demonstrated that none of the three studied mechanisms to locate data is able to satisfy S, To and N properties. Therefore combining our arguments for AP systems in the previous subsection, we proved that these three properties cannot be met simultaneously in both potentially viable distributed systems i.e., systems that guarantee strong consistency (CP) and systems that guarantee availability (AP).

\section{Limitations and Implications}\label{sec:limitation}

First of all, the presented conjecture takes in to account three fundamental mechanisms to locate data in a distributed environment, namely (1) the directory-based and (2) the computation-based methods as well as (3) the approaches based on the exploration of the network. To the best of our knowledge, these are all the mechanisms in literature to locate the data.
On the other hand, the present proof does not take into account any other mechanisms (if they ever exist) that may be potentially used for locating data. 
This can be regarded as one of the important limitations of this study. We are also aware that as new mechanisms emerge in the distributed system research, the conjecture as well as its proof may need to be revised.

One of the consequences of our result is that classical approaches cannot be used to create location protocols in a Fog computing environment because, as defined in the introduction section, the ideal Fog storage solution requires
(i) storing the data locally which would imply no  constraints on the data placement (N),
(ii) reducing the network traffic, (iii) supporting the mobility which is tolerant to the dynamicity of the network (To), and finally, which is (iv) scalable and is able to support a large number of Fog servers while being able to store a large amount of data.

%we need a system that scale (S) to a huge number of Fog sites and that is able to store a huge number of data,
%that supports mobility and therefore which is tolerant to dynamicity of the network (To)
%and finally that stores the data where needed without any constraint (N).
%As we saw in Section~\ref{sec:comparative}, no using a local database often violates the scalability property.
%Using a consistent hashing approach violates the flexibility about the constraints and
%using an exploration of the network
%This result reduces the chances to create the ideal Fog storage solution as defined in introduction,
%which provides the following properties (i) writing data locally, (ii) contains the network traffic, (iii) %supports mobility and (iv) scalability.

One of the other consequences that we  expect to come out of this result is that protocols such as \textit{Mobile IP}~\cite{rfc6275}
shall not be able to find a distributed method to store the locations of the nodes that would require all three properties at the same time. This observation also directly impacts the way we  distribute information in multiple controllers of a Software Defined Network~\cite{10.1145/2491185.2491193} which will bring forth many more implications of our result on the design of next generation communication protocols. 

\section{Conclusion and Future work}\label{sec:conclusions}

In this article, we initially reviewed different protocols for locating data fragments in a distributed environment and grouped them under three main categories. Some of these solutions rely on a local database. Others use either a computational method or are based on an exploration of the network at the data request time. We argued that no solutions from all three categories are able to meet the requirements for deployment in a distributed environment such as Fog infrastructures. In other words, these three solutions could not meet the scalability (S), the tolerance to dynamic physical network topology changes (To) without imposing constraints on the data placement or naming process (N). We finally proved our conjecture with regard to the CAP context by structuring our arguments orthogonally. We showed no protocol can satisfy the three properties at the same time for both available and partition tolerant and consistent and partition tolerant solutions. This implies that the ideal storage solution for some of the distributed environments such as Fog computing cannot satisfy the properties we expect. As future work, we plan to extend our proof for emerging data locator protocols if they ever appear in literature and look for methods to achieve the best compromise between these properties for a given distributed system application.

\vspace{-.2cm}

\bibliographystyle{unsrt} 
\bibliography{document}

\end{document}